\documentclass[a4paper,11pt]{article}
\usepackage{pos}
\usepackage{natbib}
\usepackage{xspace}
\usepackage{makecell}
\usepackage{sidecap}
\title{Dissecting the Diffuse Emission of the Galaxy with the HAWC Observatory}
 \ShortTitle{Dissecting the Diffuse Emission of the Galaxy with the HAWC Observatory}

\author*{Georg Schwefer}

\onbehalf{on behalf of the HAWC collaboration} 
\affiliation{Max-Planck-Institut für Kernphysik, Saupfercheckweg 1, 69117 Heidelberg, Germany}

\emailAdd{georg.schwefer@mpi-hd.mpg.de}

\abstract{Galactic diffuse gamma-ray emission is produced by the interaction of high-energy cosmic rays propagating through the Milky Way with interstellar gas and radiation fields. Its measurement can provide crucial insights into the acceleration and transport of cosmic rays throughout our Galaxy.

Here, we present a new analysis of the TeV Galactic diffuse gamma-ray emission using 8 years of HAWC data. This data was processed with the updated \emph{Pass 5} processing, enhancing the sensitivity and resolution of the instrument. For the analysis, we make use of \texttt{Gammapy}, an open-source package for gamma-ray astronomy, and recent models of the Galactic diffuse emission at TeV energies.

After subtracting the emission from sources using an algorithm akin to that developed for the foreseen CTAO Galactic plane survey, we find significant remaining emission throughout the Galactic plane. We show the latitudinal and longitudinal flux profiles of the emission in multiple parts of the galaxy taking into account various sources of uncertainty and compare to existing models. We find significant emission beyond 3 degree latitude, consistent in shape with the prediction for the interaction of cosmic rays with the interstellar gas.

We also demonstrate that our results are consistent with recent LHAASO results when equivalent analysis methods are used.

}

\ConferenceLogo{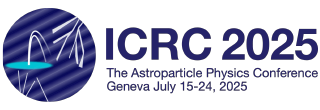}

\FullConference{39th International Cosmic Ray Conference (ICRC2025)\\
 15–24 July 2025\\
Geneva, Switzerland\\}

\begin{document}
\maketitle

\section{Introduction}

During their propagation through the Milky Way, Galactic cosmic rays (GCRs) interact with the interstellar gas and radiation fields in the Galaxy. The Galactic diffuse emission (GDE) of both gamma rays and neutrinos produced in these interactions offers unique insights into the spatial and spectral distribution of GCRs in the Milky Way (e.g.~\citep{2015ARA&A..53..199G, Tibaldo:2021viq}) and thus towards answering the question of the origin and propagation of GCRs. 
Over the last 15 years, Fermi-LAT has extensively measured the gamma-ray GDE at GeV energies (e.g.~\cite{Fermi-LAT:2016zaq, Tibaldo:2021viq}). At TeV energies, first measurements were obtained by H.E.S.S.~\cite{HESS:2014ree} a decade ago. In the past few years, the study of the TeV GDE has gained singificant new momentum with the emergence of new measurements both in neutrinos by the IceCube Observatory~\cite{doi:10.1126/science.adc9818} and in gamma rays by the Tibet-AS$\gamma$~\cite{TibetASgamma:2021tpz}, LHAASO~\cite{LHAASO:2023gne,LHAASO:2024lnz} and HAWC observatories~\cite{HAWC:2023wdq}. These measurements have 
led to a surge in modeling and interpretation efforts (e.g.~\cite{He:2025oys,DeLaTorreLuque:2025zsv,Castro:2025wgf,Schwefer:2022zly,Luque:2022buq}), debating questions such as the contribution from different classes of potential unresolved sources, the inhomogeneity of the CR spectrum in the Milky Way and the impact of different estimates of the gas distribution in the Milky Way. This underlines the importance of these measurements for a global understanding of the TeV-PeV landscape of our Galaxy.

Here, we present first results from an updated study - featuring both new and reprocessed data and enhanced analysis tools and methods - of the TeV diffuse emission of the Milky Way with the HAWC observatory. In section~\ref{sec:method}, we first describe our data and analysis methods. Our results are then presented in section~\ref{sec:results} and we summarize in~\ref{sec:summary}.

\section{Data Analysis}
\label{sec:method}

\subsection{The Dataset}

The HAWC observatory is located at an altitude of $4100\,\mathrm{m}$ in the Pico de Orizaba National Park near Puebla, Mexico. It consists of 300 water-Cherenkov detectors covering an area of $22000\,\mathrm{m}^2$.
Our dataset consists of HAWC data collected between March 2015 and January 2023 with a total lifetime of 2460 sky transits, a sizable increase compared to the 1347 transits used in the previous HAWC analysis~\cite{HAWC:2023wdq}. The events are selected and processed with the updated \emph{Pass 5} data processing~\cite{HAWC:2024plu}. It features both improved reconstruction and gamma-hadron separation, enhancing the sensitivity of the instrument in particular at large zenith angles.

The selected events are binned in two quantities: the fraction of tanks that have registered a hit, $f_{\rm hit}$~\cite{HAWC:2024plu}, and gamma-ray energy reconstructed using a neural network method~\cite{HAWC:2019xhp}. In each bin, we restrict the declination range used in the analysis based on the quality of the reconstruction in each declination band, requiring the 68 \% containment fraction of the point spread function model to be within $1.5^{\circ}$, the energy bias within $\pm100 \,\%$ of the true energy, and the energy resolution within $100 \, \%$.

To model the residual hadronic background, we use the technique described in~\cite{HAWC:2022aoz}. We mask out the Galactic plane for $|b|<10^{\circ}$ in the background-modeling procedure to avoid any diffuse emission being absorbed in the background. To minimize the impact of systematic uncertainties associated with the CR background, arising for example from the anisotropy in the arrival direction of cosmic rays on earth, we exclude the first four  $f_{\rm hit}$ bins~\cite{HAWC:2024plu}. This condition raises the low-energy threshold of the analysis to $1\,\mathrm{TeV}$.

\subsection{Analysis methods}
\label{sec:analysis_methods}

To perform the analysis, we use \texttt{Gammapy}, a \texttt{Python} package for gamma-ray astronomy \cite{Gammapy:2023gvb}, version 1.3 \cite{acero_2025_14760974}. For the results presented here, we use two separate analysis methods.

For the first and main analysis, we fit and subsequently subtract a model for the source emission from our data and then analyze the residual diffuse emission. This model for the source emission is obtained using a workflow following that developed for the foreseen CTAO Galactic Plane Survey (for details see appendix D of~\cite{CTAConsortium:2023tdz}).

This source modeling workflow was simplified and adapted for its application to HAWC data.
Significance maps of the excesses above the CR background were computed for larger correlation radii of $0.2^{\circ}$ and $0.4^{\circ}$. Candidate objects detected as local maxima above 3 sigma in these maps were imposed to have a minimal angular separation of 0.2$^{\circ}$ or a radius difference larger than 0.5$^{\circ}$, otherwise they were merged into a single object. The fitting step was simplified to consider only a generalized Gaussian as spatial model and a log-parabola spectral model for each object. In the end, this algorithm results in a list of candidate sources with their best-fit parameters and detection significance.

Having subtracted the sources, we then make use of state-of-the-art models to fit the remaining GDE flux. In particular, we use the hadronic part of the CRINGE model developed in~\cite{Schwefer:2022zly} with a rescaled value of $X_{\rm CO}$ following~\cite{CTAConsortium:2023tdz}, the IEM-varmin-rescaled model from~\cite{CTAConsortium:2023tdz} and a Uniform CR model. All three models are based on a fit to the local spectrum of Galactic CRs, but differ in their prediction in other parts of the Galaxy: The Uniform CR model assumes the local values of density and spectrum of CRs as used in the CRINGE model to be uniformly present everywhere in the Miky Way. In this model then, the diffuse emission intensity from a given line-of-sight is simply proportional to the gas column density. In the CRINGE model, the distribution of CRs arises from an imhomogeneous distribution of CR sources and the propagation of CRs in the Milky Way with a homogeneous diffusion coefficient~\cite{Schwefer:2022zly}. The spectrum is homogeneous. The IEM-varmin-rescaled model differs from it in that it assumes an inhomogeneous scaling of the diffusion coefficient, leading to a spectral hardening of CRs towards the Galactic center. Additionally, the distribution of CRs as a function of galactocentric radius assumed for this model is adjusted to match Fermi-LAT observations~\cite{CTAConsortium:2023tdz,Fermi-LAT:2016zaq}.

\begin{figure*}[!ht]
\centering
\includegraphics[trim={0cm 0.7cm 0cm 0.3cm},clip,width=0.99\textwidth]{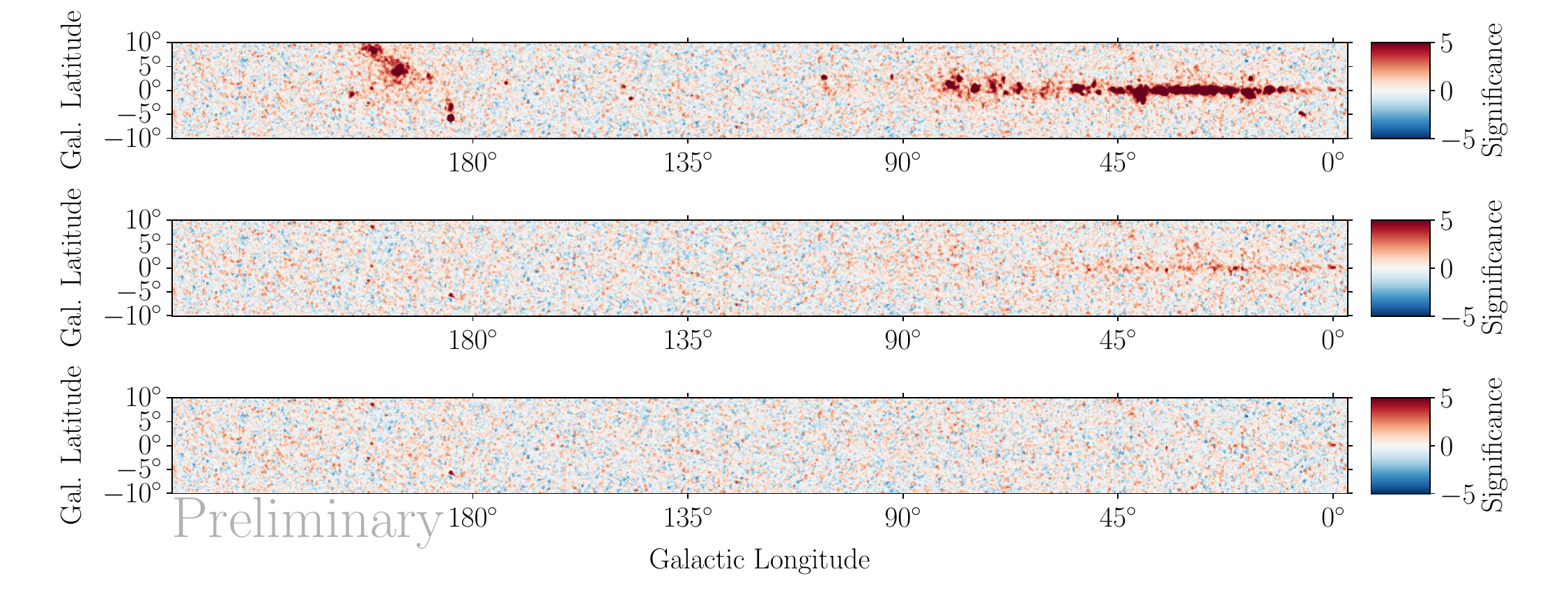}
\caption{
Significance maps of our dataset with a correlation radius of $0.4^{\circ}$ between $-10^{\circ}<b<10^{\circ}$ and $0^{\circ}<l<240^{\circ}$. In the upper panel, we show the raw significance map including all sources and diffuse emission. The middle panel shows the residual significance after subtraction of the source model. The lower panel shows the residual significance after further subtraction of the CRINGE diffuse emission model, multiplied in each longitude bin by the best-fit normalization as used for the longitudinal profile in Figure~\ref{fig:lon_profile}.  
}
\label{fig:significance_maps}
\end{figure*}

The significance map of the dataset and the source subtraction process are visualized in Figure~\ref{fig:significance_maps}. The impact of the source subtraction between the upper and middle panel as well as the subtraction of the diffuse emission from the middle to the lower panel is clearly visible in the maps. The latter improves the quality of the fit particularly towards the Galactic center. Small residual overfluctuations remain near the Crab Nebula and Geminga. These are a consequence of the difficulty of perfectly fitting these very high significance sources with the geometric spatial models used in the source modeling algorithm. On large scales however, these plots validate our approach of fitting the emission of the Galactic plane with the methods described above. 

 The goal of the second analysis is to compare as closely as possible our results with the existing studies of the Galactic diffuse emission by the LHAASO collaboration~\cite{LHAASO:2023gne,LHAASO:2024lnz}. To this end, we attempt to reproduce the analysis methods used therein. We split our data in the same "inner" ($|b|<5^{\circ}$, $15^{\circ}<l<125^{\circ}$) and "outer" ($|b|<5^{\circ}$, $125^{\circ}<l<235^{\circ}$) analysis regions and use the mask provided in the supplementary material of~\cite{LHAASO:2023gne} to exclude known sources. We do not subtract any sources model here. To fit the parameters of our spectral model, we use the same Poisson likelihood summing events in each energy bin as described in~\cite{LHAASO:2023gne}. The result of this analysis is presented in section~\ref{sec:lhaaso_results}.

\section{Results}
\label{sec:results}

\subsection{Emission profiles}

\begin{figure*}[!ht]
\centering
\includegraphics[width=0.99\textwidth]{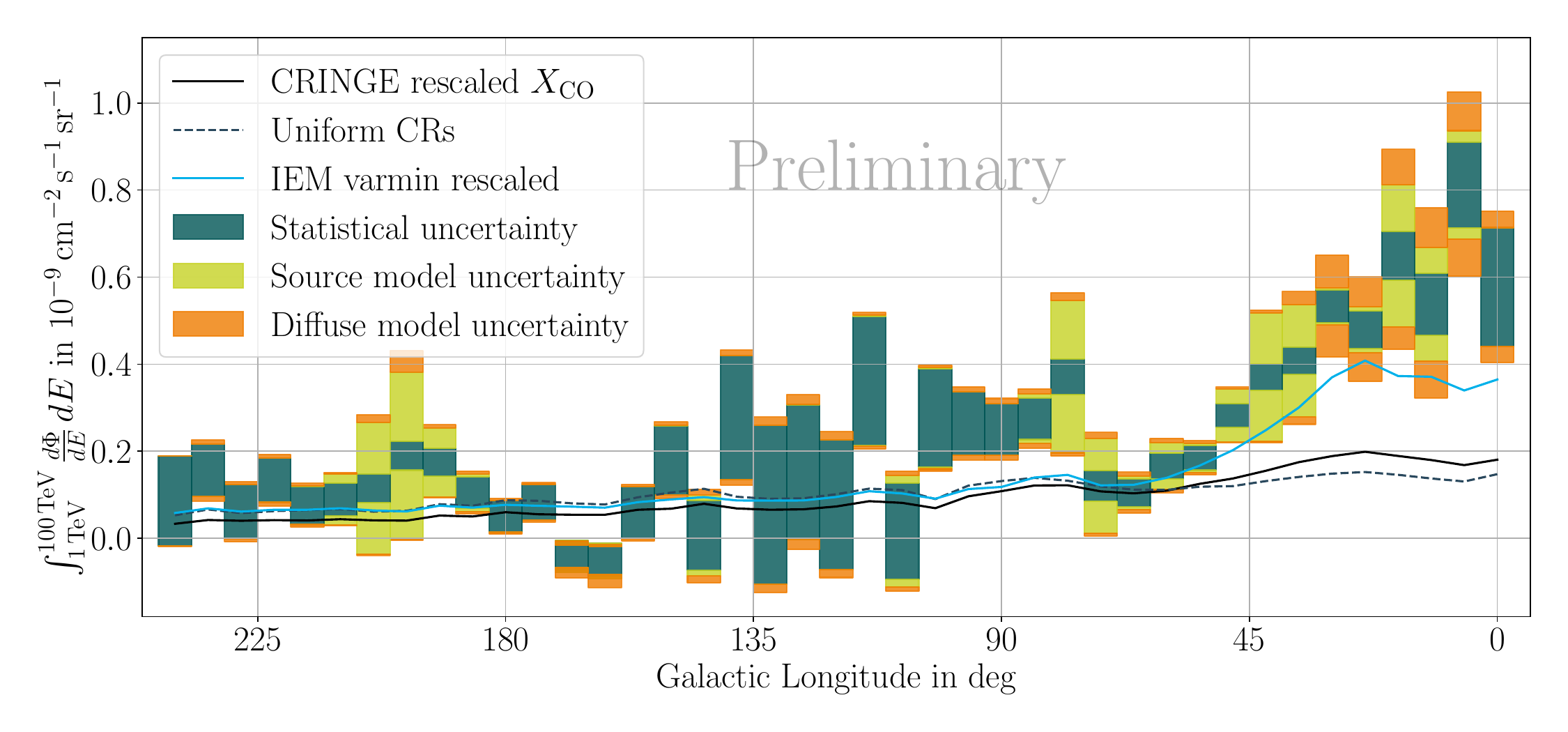}

\caption{
Longitudinal profile in integrated flux between 1 and 100 TeV of the residual Galactic emission after subtraction of the source model between $-10^{\circ}<b<10^{\circ}$ and $0^{\circ}<l<240^{\circ}$. We show three different uncertainties associated with the measurement as well as three different models for the GDE of the Milky Way described in the text.
}
\label{fig:lon_profile}
\end{figure*}

\begin{SCfigure}
\centering
\includegraphics[width=0.7\textwidth]{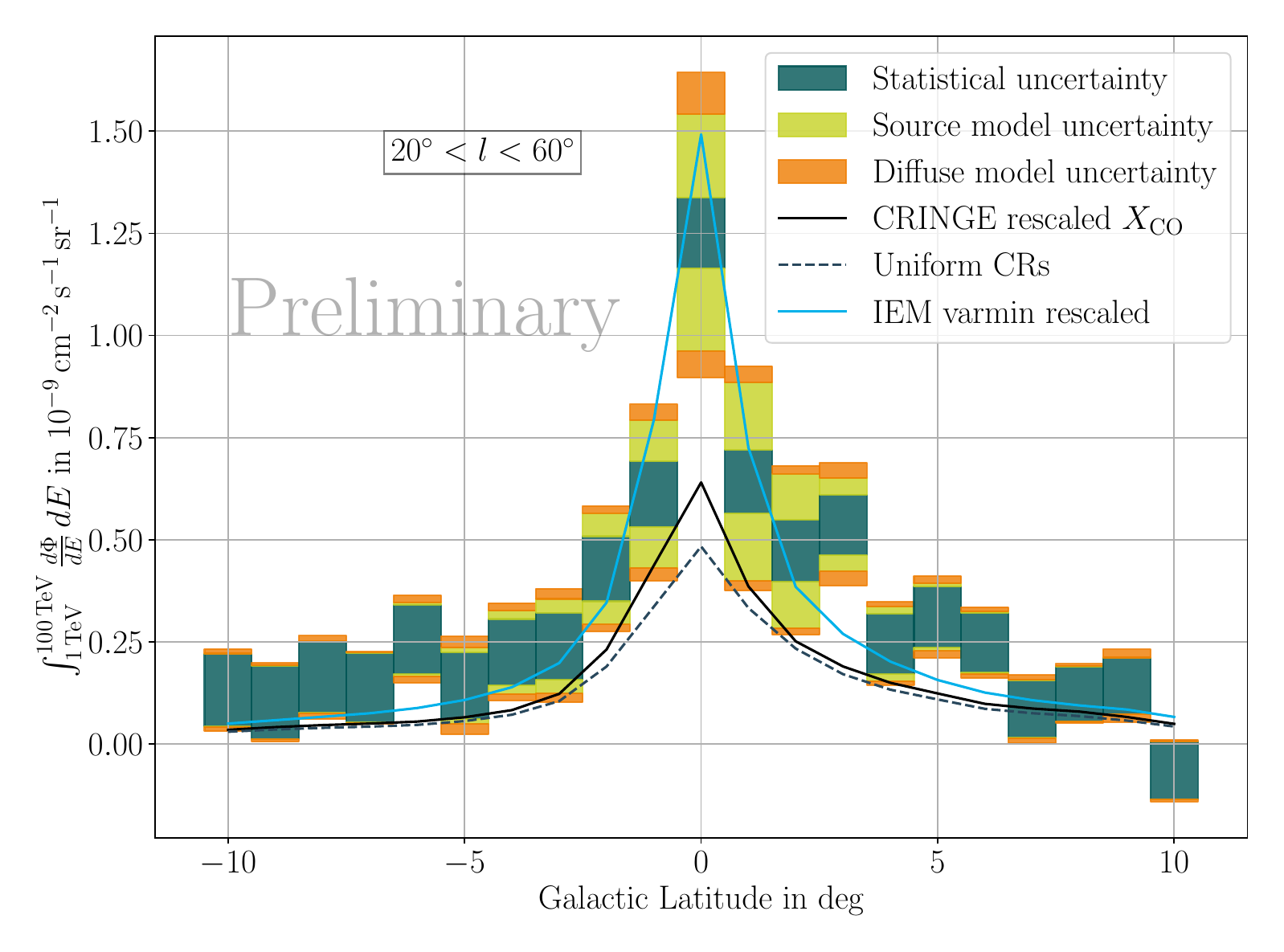}

\caption{
Latitudinal profile of the residual Galactic emission after subtraction of the source model between \mbox{$-10^{\circ}<b<10^{\circ}$} and \mbox{$20^{\circ}<l<60^{\circ}$}. We show three different uncertainties associated with the measurement as well as three different models for the GDE of the Milky Way described in the text.\vspace{2.5cm}
}
\label{fig:lat_profile}
\end{SCfigure}

In Figures~\ref{fig:lon_profile} and~\ref{fig:lat_profile}, we show the longitudinal and latitudinal distributions in integrated flux between 1 and 100 TeV of the residual emission after subtraction of the source model in the Galactic plane.
The profiles show three different uncertainties. The first is the statistical uncertainty, the second is the source model uncertainty and the third is the uncertainty associated with the choice of diffuse emission model.
We obtain these from in total six individual fits, assuming two different source models and the three different diffuse emission models mentioned in section~\ref{sec:analysis_methods}.

The two source models are obtained by assuming two different thresholds for source selection. In one case, all possible sources picked up by the source detection algorithm are included in the model. In the other case, only sources with a detection significance of at least $4\sigma$ or $3\sigma$ and an association with a source in a different VHE catalog are included. The difference between the two thresholds mainly includes large, low-surface brightness objects for which there could be confusion between source and diffuse emission.

For each region the errorband is centered on the mean of the six fits. The statistical error is the mean statistical error of the six fits. The source model uncertainty is given by the difference between the mean fit result for each of the two source models assumed. The upper and lower diffuse emission model uncertainties are given by the standard deviation of the fits with the three diffuse emission model for either of the source models. 

Regarding the trend of the mean values of the longitudinal profile, we can see a general increase in the integrated diffuse flux towards the Galactic center, as predicted by all models. The rise is particularly pronounced for $l<50^{\circ}$, similar to what is seen at lower energies by Fermi-LAT~\cite{Tibaldo:2021viq}. This could be an indication that the gamma-ray emissivity at low Galactocentric radii is increased compared to the local value and predictions by conventional models such as CRINGE. There are a number of possible origins of this, including a distribution of CR sources that is more peaked towards the Galactic center than assumed in the models, a radial gradient in the CR accelerator properties or a non-standard cosmic-ray propagation scenario~\cite{DeLaTorreLuque:2025zsv} such as the one modeled in the IEM-varmin-rescaled model. Alternatively, it could also be caused by a population of unresolved sub-threshold sources (e.g.~\cite{LHAASO:2024lnz,Pagliaroli:2023naa}).
The latitudinal profile of the emission as expected is peaked in the Galactic plane. It also shows extended, clearly non-gaussian tails all the way out to $10^{\circ}$ latitude. These are expected from emission associated with the gas in the Milky Way, as the atomic gas extends up to hundreds of parsecs out of the Galactic plane~\cite{2009ARA&A..47...27K}.
Consistent with the longitudinal profile in the region considered, the overall emission is enhanced with respect to the prediction by conventional models such as CRINGE. Interestingly, this enhancement is present at all latitudes.

Regarding the different uncertainties to the measurement, we note the following observations:
The statistical uncertainty is to a large extent proportional to the exposure at a given Galactic longitude, which itself is proportional to the culminating zenith angle of this longitude for HAWC. This also explains why the statistical uncertainty is almost constant in all regions in the latitudinal profile. In the longitudinal profile, we observe three regions with large source model uncertainties: The first two are near Geminga and Monogem at $l\approx200^{\circ}$ and near the Cygnus region around $l\approx75^{\circ}$. This makes sense as these regions feature large, low-surface brightness, extended sources~\cite{LHAASO:2023uhj,HAWC:2024scl} that can be confused with the diffuse emission and are difficult to model. 
The third region with generally larger source model uncertainty in the longitudinal profile is towards the inner galaxy, where HAWC has large exposure and sees a high density of sources (see Figure~\ref{fig:significance_maps}). It is logical that because of overlap and confusion of sources there is a larger absolute uncertainty associated with the source model.
The last point is also visible in the latitudinal profile: The source model uncertainty is largest in the Galactic plane, precisely where we find the highest density of sources. Note that this means in reverse that the presence of the tails of the emission above $|b|>3^{\circ}$ is independent of the source model assumed.
Finally, we can see in Figures~\ref{fig:lon_profile} and~\ref{fig:lat_profile} that the diffuse emission model uncertainty becomes largest towards the Galactic center and in the Galactic plane. This is also expected as this is precisely where the differences between the models tested, both in terms of energy spectrum and morphology, become largest.

\subsection{Consistency with LHAASO results}
\label{sec:lhaaso_results}

Here, we present the results from the analysis trying to reproduce as faithfully as possible the LHAASO analysis from~\cite{LHAASO:2023gne,LHAASO:2024lnz} as described in section~\ref{sec:analysis_methods}.

\begin{figure*}[!ht]
\centering
\includegraphics[width=0.49\textwidth]{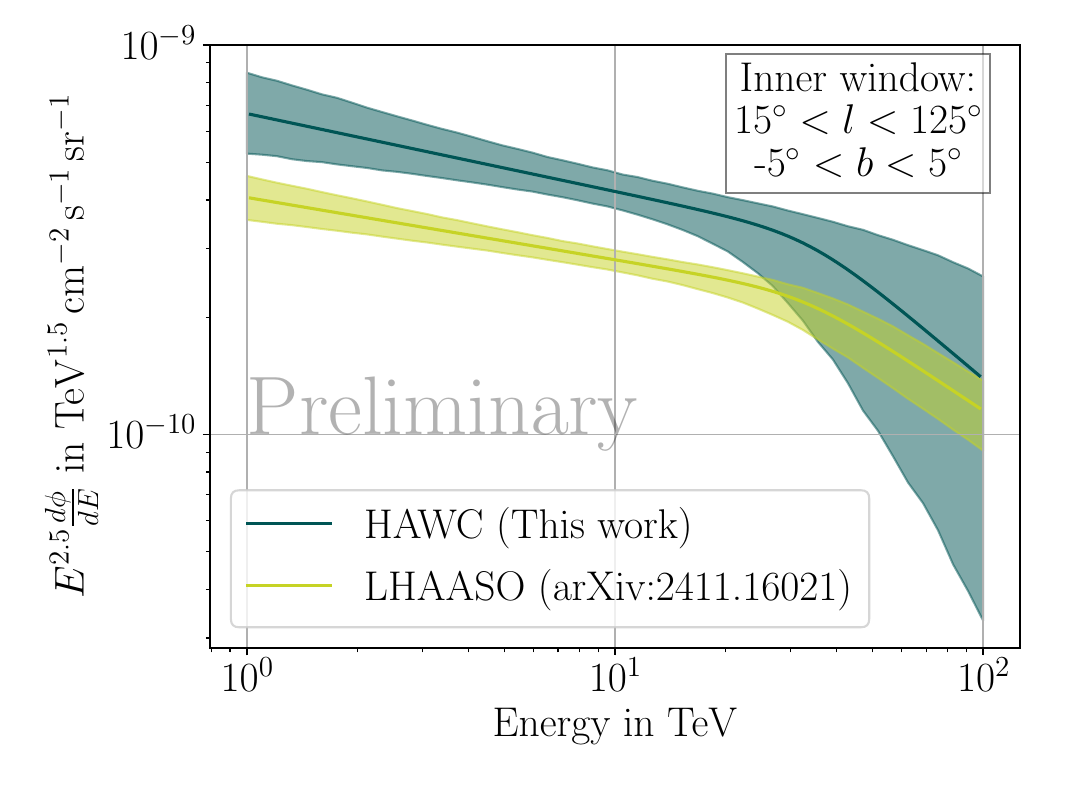}
\includegraphics[width=0.49\textwidth]{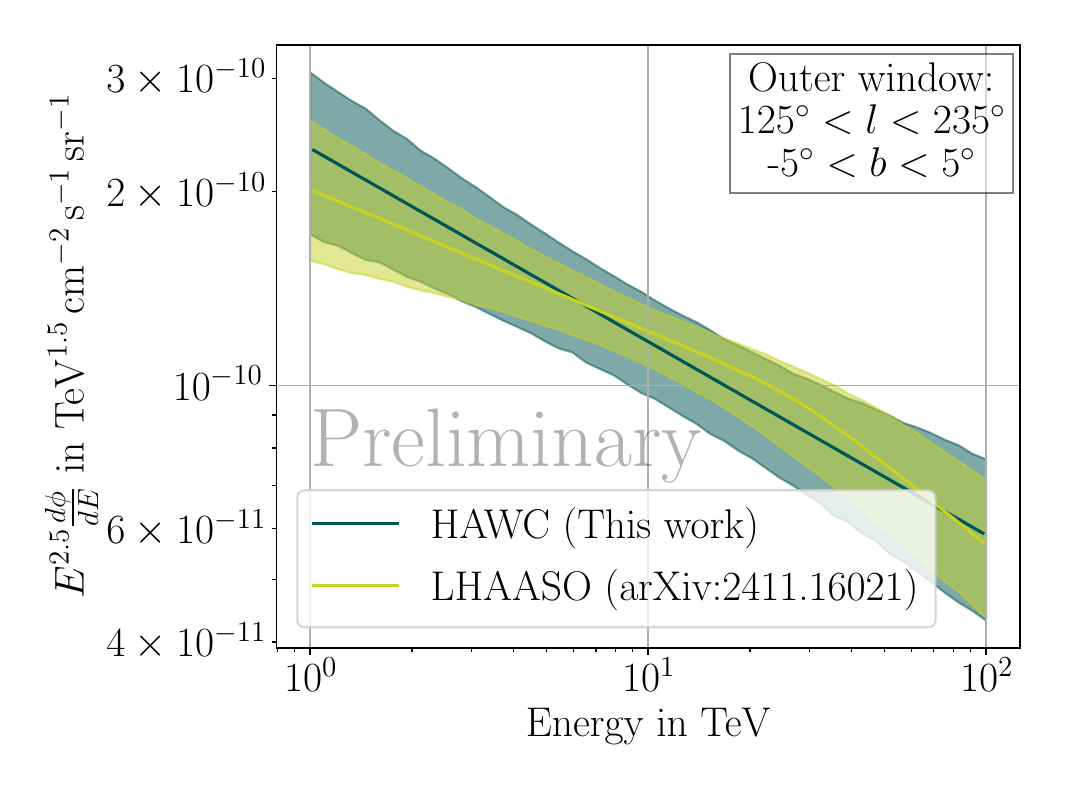}
\caption{
Spectrum of the Galactic diffuse gamma-ray emission in the inner (left) and outer (right) Galactic window obtained from the analysis following the LHAASO methods. We also show the corresponding LHAASO results from~\cite{LHAASO:2024lnz}. The respective best-fit parameters and uncertainties are given in Tables~\ref{tab:lhaaso_comp_inner} and \ref{tab:lhaaso_comp_outer}.
}
\label{fig:lhaaso_comp_plots}
\end{figure*}

In the inner window, we fit a broken power-law with the same functional form as given in~\cite{LHAASO:2024lnz} and can therefore do a one-to-one parameter comparison to the LHAASO result from that study. The best-fit parameters and uncertainties from this study and the corresponding LHAASO results from~\cite{LHAASO:2024lnz} are given in Table~\ref{tab:lhaaso_comp_inner}. An illustration of the corresponding spectra is shown in the left panel of Figure~\ref{fig:lhaaso_comp_plots}. We find an overall very similar spectral shape to the LHAASO result, with consistent values for the spectral indices below and above the spectral break. The best-fit energy of this break is also very close to the LHAASO best fit, albeit with large uncertainties. At lower energies, we fit a flux normalization that is about $50\%$ larger than the one measured by LHAASO. A possible explanation for this is the difference in the mask of the Galactic sources between the one from~\cite{LHAASO:2023gne} that is publicly available and used here and the one from~\cite{LHAASO:2024lnz}. As the former masks out slightly less of the Galactic plane than the latter in the inner window, an increased flux normalization is expected in our result.

In the outer window, we can not constrain and find any evidence for a spectral break. Therefore, we only fit a single power-law spectral model. Note that the spectral break in the LHAASO result is also not statistically significant~\cite{LHAASO:2024lnz}. The best-fit parameters and uncertainties from this study and the corresponding LHAASO results from~\cite{LHAASO:2024lnz} are given in Table~\ref{tab:lhaaso_comp_outer}. An illustration of the corresponding spectra is shown in the right panel of Figure~\ref{fig:lhaaso_comp_plots}. We find very good consistency with the LHAASO result both in terms of spectrum and flux normalization. This is particularly interesting because here, the mask covers less of the Galactic plane than in the inner window and the difference between the mask from~\cite{LHAASO:2023gne} used in this work and the mask from~\cite{LHAASO:2024lnz} is much smaller.

\begin{table}

\centering
\caption{Best-fit parameters and uncertainties from fit in the inner Galactic window and comparison to the LHAASO result~\cite{LHAASO:2024lnz}. Both this work and the LHAASO study use the broken power-law model from~\cite{LHAASO:2024lnz}.}
\begin{tabular}{c|c|c|c|c}
&\makecell{$d\phi/dE$@ $10\,\mathrm{TeV}$\\ in $\mathrm{TeV}^{-1}\,\mathrm{cm}^{-2}\,\mathrm{s}^{-1}\,\mathrm{sr}^{-1}$}& $\gamma_1$& $\gamma_2$ & $E_{\rm br}$ in TeV \\ 
\hline HAWC (This work) & $(13.3^{+1.7}_{-1.0})\times10^{-13}$ & $2.70^{+0.09}_{-0.10}$ & $3.3^{+1.4}_{-0.3}$& $34^{+62}_{-18}$  \\
\hline LHAASO~\cite{LHAASO:2024lnz} & $(8.88\pm0.53)\times10^{-13}$ & $2.66 \pm 0.05$ & $3.13 \pm 0.08$ & $32.84\pm11.16$  \\
\end{tabular}

\label{tab:lhaaso_comp_inner}
\end{table}

\begin{table}

\centering
\caption{Best-fit parameters and uncertainties from fit in the outer Galactic window and comparison to the LHAASO result~\cite{LHAASO:2024lnz}. As we can not find any evidence for a spectral break, we fit only a power-law model.}
\begin{tabular}{c|c|c|c|c}

&\makecell{$d\phi/dE$@ $10\,\mathrm{TeV}$\\ in $\mathrm{TeV}^{-1}\,\mathrm{cm}^{-2}\,\mathrm{s}^{-1}\,\mathrm{sr}^{-1}$}& $\gamma_1$& $\gamma_2$ & $E_{\rm br}$ in TeV \\ 
\hline HAWC (This work) & $(3.70\pm0.65)\times10^{-13}$ & \multicolumn{2}{c|}{$2.8\pm 0.1$}& $-$  \\
\hline LHAASO~\cite{LHAASO:2024lnz} & $(3.84\pm0.37)\times10^{-13}$ & $2.72 \pm 0.1$ & $2.92 \pm 0.1$ & $27.86\pm22.49$  \\
\end{tabular}

\label{tab:lhaaso_comp_outer}
\end{table}

\section{Summary and Conclusions}
\label{sec:summary}

We have presented an updated and expanded analysis of the TeV Galactic diffuse gamma-ray emission using 8 years of reprocessed data from the HAWC Observatory and new analysis tools.

First, we use an approach in which we subtract a fitted source model following the algorithm developed for the CTAO Galactic plane survey~\cite{CTAConsortium:2023tdz}. Then, we use templates based on state-of-the-art models to infer the longitudinal and latitudinal profiles of the residual diffuse emission in the Galaxy. In this process, we also consider systematic uncertainties arising from the source subtraction and the choice of template. The shape of the resulting emission profiles is largely consistent with the model predictions.
Notably, the diffuse emission has non-gaussian tails beyond $|b|>3^{\circ}$ as expected from emission correlated with the Galactic gas. 
Towards the Galactic center, we see an increase with respect to the conventional model predictions such as CRINGE. This could be due to a number of factors, including a different-than-assumed source distribution, non-conventional propagation scenarios or a population of unresolved sources.

In a second analysis, we aim to reproduce as closely as possible the analysis by the LHAASO collaboration presented in~\cite{LHAASO:2023gne,LHAASO:2024lnz} and mask out sources instead of fitting and subtracting them. We find largely consistent results with LHAASO in both the inner and outer Galactic window, with a $50\%$ larger flux normalization in the inner window that can at least partially be explained by a difference in the source mask used.

We now aim to use the first analysis method to further dissect the diffuse gamma-ray emission of the Galaxy. This includes for example looking in more detail at the energy spectrum of the emission in different parts of the galaxy. Another possibility is to split the emission into various components, such as those originating from different Galactocentric radii and those associated with the molecular and atomic gas in the Milky Way as done at GeV energies with Fermi-LAT~\cite{Fermi-LAT:2016zaq}. This will help to further deepen our understanding of the processes in the Galaxy that shape the TeV-PeV landscape of the Milky Way.

\bibliographystyle{JHEP_v2}
\small
\bibliography{library_new}

\clearpage
\section*{Full Author List: \ HAWC Collaboration}
\scriptsize
\noindent

\vskip2cm
\noindent

R. Alfaro$^{1}$,
C. Alvarez$^{2}$,
A. Andrés$^{3}$,
E. Anita-Rangel$^{3}$,
M. Araya$^{4}$,
J.C. Arteaga-Velázquez$^{5}$,
D. Avila Rojas$^{3}$,
H.A. Ayala Solares$^{6}$,
R. Babu$^{7}$,
P. Bangale$^{8}$,
E. Belmont-Moreno$^{1}$,
A. Bernal$^{3}$,
K.S. Caballero-Mora$^{2}$,
T. Capistrán$^{9}$,
A. Carramiñana$^{10}$,
F. Carreón$^{3}$,
S. Casanova$^{11}$,
S. Coutiño de León$^{12}$,
E. De la Fuente$^{13}$,
D. Depaoli$^{14}$,
P. Desiati$^{12}$,
N. Di Lalla$^{15}$,
R. Diaz Hernandez$^{10}$,
B.L. Dingus$^{16}$,
M.A. DuVernois$^{12}$,
J.C. Díaz-Vélez$^{12}$,
K. Engel$^{17}$,
T. Ergin$^{7}$,
C. Espinoza$^{1}$,
K. Fang$^{12}$,
N. Fraija$^{3}$,
S. Fraija$^{3}$,
J.A. García-González$^{18}$,
F. Garfias$^{3}$,
N. Ghosh$^{19}$,
A. Gonzalez Muñoz$^{1}$,
M.M. González$^{3}$,
J.A. Goodman$^{17}$,
S. Groetsch$^{19}$,
J. Gyeong$^{20}$,
J.P. Harding$^{16}$,
S. Hernández-Cadena$^{21}$,
I. Herzog$^{7}$,
D. Huang$^{17}$,
P. Hüntemeyer$^{19}$,
A. Iriarte$^{3}$,
S. Kaufmann$^{22}$,
D. Kieda$^{23}$,
K. Leavitt$^{19}$,
H. León Vargas$^{1}$,
J.T. Linnemann$^{7}$,
A.L. Longinotti$^{3}$,
G. Luis-Raya$^{22}$,
K. Malone$^{16}$,
O. Martinez$^{24}$,
J. Martínez-Castro$^{25}$,
H. Martínez-Huerta$^{30}$,
J.A. Matthews$^{26}$,
P. Miranda-Romagnoli$^{27}$,
P.E. Mirón-Enriquez$^{3}$,
J.A. Montes$^{3}$,
J.A. Morales-Soto$^{5}$,
M. Mostafá$^{8}$,
M. Najafi$^{19}$,
L. Nellen$^{28}$,
M.U. Nisa$^{7}$,
N. Omodei$^{15}$,
E. Ponce$^{24}$,
Y. Pérez Araujo$^{1}$,
E.G. Pérez-Pérez$^{22}$,
Q. Remy$^{14}$,
C.D. Rho$^{20}$,
D. Rosa-González$^{10}$,
M. Roth$^{16}$,
H. Salazar$^{24}$,
D. Salazar-Gallegos$^{7}$,
A. Sandoval$^{1}$,
M. Schneider$^{1}$,
G. Schwefer$^{14}$,
J. Serna-Franco$^{1}$,
A.J. Smith$^{17}$
Y. Son$^{29}$,
R.W. Springer$^{23}$,
O. Tibolla$^{22}$,
K. Tollefson$^{7}$,
I. Torres$^{10}$,
R. Torres-Escobedo$^{21}$,
R. Turner$^{19}$,
E. Varela$^{24}$,
L. Villaseñor$^{24}$,
X. Wang$^{19}$,
Z. Wang$^{17}$,
I.J. Watson$^{29}$,
H. Wu$^{12}$,
S. Yu$^{6}$,
S. Yun-Cárcamo$^{17}$,
H. Zhou$^{21}$,

\vskip2cm
\noindent

$^{1}$Instituto de F\'{i}sica, Universidad Nacional Autónoma de México, Ciudad de Mexico, Mexico,
$^{2}$Universidad Autónoma de Chiapas, Tuxtla Gutiérrez, Chiapas, México,
$^{3}$Instituto de Astronom\'{i}a, Universidad Nacional Autónoma de México, Ciudad de Mexico, Mexico,
$^{4}$Universidad de Costa Rica, San José 2060, Costa Rica,
$^{5}$Universidad Michoacana de San Nicolás de Hidalgo, Morelia, Mexico,
$^{6}$Department of Physics, Pennsylvania State University, University Park, PA, USA,
$^{7}$Department of Physics and Astronomy, Michigan State University, East Lansing, MI, USA,
$^{8}$Temple University, Department of Physics, 1925 N. 12th Street, Philadelphia, PA 19122, USA,
$^{9}$Universita degli Studi di Torino, I-10125 Torino, Italy,
$^{10}$Instituto Nacional de Astrof\'{i}sica, Óptica y Electrónica, Puebla, Mexico,
$^{11}$Institute of Nuclear Physics Polish Academy of Sciences, PL-31342 11, Krakow, Poland,
$^{12}$Dept. of Physics and Wisconsin IceCube Particle Astrophysics Center, University of Wisconsin{\textemdash}Madison, Madison, WI, USA,
$^{13}$Departamento de F\'{i}sica, Centro Universitario de Ciencias Exactase Ingenierias, Universidad de Guadalajara, Guadalajara, Mexico, 
$^{14}$Max-Planck Institute for Nuclear Physics, 69117 Heidelberg, Germany,
$^{15}$Department of Physics, Stanford University: Stanford, CA 94305–4060, USA,
$^{16}$Los Alamos National Laboratory, Los Alamos, NM, USA,
$^{17}$Department of Physics, University of Maryland, College Park, MD, USA,
$^{18}$Tecnologico de Monterrey, Escuela de Ingenier\'{i}a y Ciencias, Ave. Eugenio Garza Sada 2501, Monterrey, N.L., Mexico, 64849,
$^{19}$Department of Physics, Michigan Technological University, Houghton, MI, USA,
$^{20}$Department of Physics, Sungkyunkwan University, Suwon 16419, South Korea,
$^{21}$Tsung-Dao Lee Institute \& School of Physics and Astronomy, Shanghai Jiao Tong University, 800 Dongchuan Rd, Shanghai, SH 200240, China,
$^{22}$Universidad Politecnica de Pachuca, Pachuca, Hgo, Mexico,
$^{23}$Department of Physics and Astronomy, University of Utah, Salt Lake City, UT, USA, 
$^{24}$Facultad de Ciencias F\'{i}sico Matemáticas, Benemérita Universidad Autónoma de Puebla, Puebla, Mexico, 
$^{25}$Centro de Investigaci\'on en Computaci\'on, Instituto Polit\'ecnico Nacional, M\'exico City, M\'exico,
$^{26}$Dept of Physics and Astronomy, University of New Mexico, Albuquerque, NM, USA,
$^{27}$Universidad Autónoma del Estado de Hidalgo, Pachuca, Mexico,
$^{28}$Instituto de Ciencias Nucleares, Universidad Nacional Autónoma de Mexico, Ciudad de Mexico, Mexico, 
$^{29}$University of Seoul, Seoul, Rep. of Korea,
$^{30}$Departamento de Física y Matemáticas, Universidad de Monterrey, Av.~Morones Prieto 4500, 66238, San Pedro Garza Garc\'ia NL, M\'exico

\end{document}